\documentclass[3p,times,twocolumn]{elsarticle}

\usepackage{ecrc}
\usepackage{amsmath}

\usepackage{xcolor}
\definecolor{lcolor}{rgb}{0.5,0,0}
\definecolor{citcolor}{rgb}{0,0.3,0.0}

\usepackage[breaklinks,colorlinks,urlcolor=blue,citecolor=citcolor,linkcolor=lcolor]{hyperref}

\volume{00}

\firstpage{1}

\journalname{Nuclear and Particle Physics Proceedings}

\runauth{}

\jid{nppp}

\jnltitlelogo{Nuclear and Particle Physics Proceedings}

\usepackage{amssymb}

\usepackage[figuresright]{rotating}

\newcommand{\as}{\alpha_{\mathrm{s}}}
\newcommand{\nc}{{N_\mathrm{c}}}
\newcommand{\nf}{{n_\mathrm{f}}}
\newcommand{\tr}{\, \mathrm{Tr} \, }
\newcommand{\ud}{\, \mathrm{d}}
\newcommand{\lqcd}{\Lambda_{\mathrm{QCD}}}
\newcommand{\qso}{Q_\mathrm{s,0}}
\newcommand{\qs}{Q_\mathrm{s}}
\newcommand{\nr}[1]{(\ref{#1})}
\newcommand{\eq}{Eq.~}

\begin{document}

\begin{frontmatter}



\dochead{}

\title{Including resummation in the NLO BK equation}


\author{}

\address{}

\author[jyfl,hip]{T. Lappi}
\address[jyfl]{
Department of Physics, University of Jyv\"askyl\"a %
 P.O. Box 35, 40014 University of Jyv\"askyl\"a, Finland
}

\address[hip]{
Helsinki Institute of Physics, P.O. Box 64, 00014 University of Helsinki,
Finland
}
\author[bnl]{H. M\"antysaari}
\address[bnl]{
Physics Department, Brookhaven National Laboratory, Upton, NY 11973, USA
}

\begin{abstract}
We include a resummation of large transverse momentum logarithms in the next-to-leading order (NLO) Balitsky-Kovchegov equation. The resummed evolution equation is shown to be stable, the evolution speed being significantly reduced by NLO corrections. The contributions from NLO terms that are not enhanced by large logarithms are found to be numerically important close to phenomenologically relevant initial conditions. We numerically determine the value for the constant in the resummed logarithm that includes a maximal part of the full NLO terms in the resummation.
\end{abstract}


\end{frontmatter}

\section{Introduction}

Many calculations in the CGC~\cite{Gelis:2010nm} framework for describing high energy collisions in QCD are currently advancing to NLO accuracy in the weak coupling QCD perturbative expansion. This is the case for the Balitsky-Kovchegov~\cite{Balitsky:2008zza} and JIMWLK~\cite{Balitsky:2013fea,Kovner:2013ona} renormalization group equations as well as for cross sections of specific processes such as inclusive DIS~\cite{Balitsky:2010ze,Beuf:2011xd} and forward single inclusive particle production in proton-nucleus collisions.  
Armed with these results, one would like to use them for phenomenology and evaluate the cross sections. This talk describes one part of this program, the numerical solution of the NLO BK equation both in its original form, following~\cite{Lappi:2015fma} and subsequently~\cite{Lappi:2016fmu}, including a resummation of large double and single collinear logarithms following the procedure developed in~\cite{Iancu:2015vea,Iancu:2015joa}.

After writing down the equation, we will first demonstrate the instability problem of the original unresummed equation, which manifests itself as a $\ln r$-divergence in the evolution speed of small dipoles. We then discuss our implementation of the resummation of the problematic logarithms. Finally we analyze numerically both the equation containing only the resummation and the full equation including the remaining finite terms.

\section{The NLO BK equation}
The large $\nc$-limit of the NLO BK equation as derived in \cite{Balitsky:2008zza} can be written as
\begin{multline}
\label{eq:nlobk}
\partial_y S(r) = \frac{\as \nc}{2\pi^2} {\boldsymbol {K_1 }} \otimes [S(X)S(Y)-S(r)] 
\\
+ \frac{\as^2 \nf \nc}{8\pi^4} {\boldsymbol{K_f}} \otimes S(Y)[S(X')-S(X)]
\\
+ \frac{\as^2 \nc^2}{8\pi^4} \boldsymbol{K_2} \otimes [S(X)S(z-z')S(Y')-S(X)S(Y)].
\end{multline}
The equation describes the dependence on rapidity $y= \ln s$ of the dipole operator, i.e. a correlator of two fundamental representation Wilson lines in the color field of the target:
\begin{equation}
S(x-y) \equiv 1 - N(x-y) \equiv \frac{1}{\nc} \left< \tr U^\dag(x) U(y)\right>.
\end{equation}
The Wilson lines in the target are taken at coordinates $x,y,z,z'$ in the transverse plane, with $\otimes$ denoting an integral over $z$ or, depending on the term, $z$ and $z'$. We denote $r=x-y$, $X=x-z$, $Y=y-z$, $X'=x-z'$ and $Y'=y-z'$.
Here we have also closed the equation by using the large-$\nc$ mean-field approximation that replaces expectation values of products of dipoles by the product of expectation values.
The kernels 
\begin{align}
\nonumber
K_1 &= {\frac{r^2}{X^2Y^2} \bigg[ 1}+
\frac{\as\nc }{4\pi} \bigg( 
{ \frac{\beta}{\nc} \ln r^2\mu^2 - \frac{\beta}{\nc} \frac{X^2-Y^2}{r^2} \ln \frac{X^2}{Y^2} } 
\\ &  + \frac{67}{9} - \frac{\pi^2}{3} - \frac{10}{9} \frac{\nf}{\nc}
 - {2 \ln \frac{X^2}{r^2} \ln \frac{Y^2}{r^2}}
 \bigg) \bigg]
\\
K_2 &= -\frac{2}{(z-z')^4} + \bigg[ \frac{X^2 Y'^2 + X'^2Y^2 - 4r^2(z-z')^2}{(z-z')^4(X^2Y'^2 - X'^2Y^2)} 
\\
\nonumber
& 
+ \frac{r^4}{X^2Y'^2(X^2Y'^2 - X'^2Y^2)} + \frac{r^2}{X^2Y'^2(z-z')^2} \bigg]
{ \ln \frac{X^2Y'^2}{X'^2Y^2}} \nonumber
\\
 K_f &= \frac{2}{(z-z')^4}  
	- \frac{X'^2Y^2 + Y'^2 X^2 - r^2 (z-z')^2}{(z-z')^4(X^2Y'^2 - X'^2Y^2)}{  \ln \frac{X^2Y'^2}{X'^2Y^2} }
\end{align}
include notably of the leading order part (first term in $K_1$), running coupling terms (proportional to the beta function coefficient $\beta$), and logarithms of both conformal (the ones that vanish when $r=0$, i.e. $X=Y$ and $X'=Y'$) and nonconformal (in $K_1$) ratios of dipole sizes. We absorb all the terms involving the beta function coefficient $\beta$ into a running coupling constant, eliminating the renormalization scale $\mu$. We do this via the   ``Balitsky'' running coupling for the LO term, and a parent dipole prescription for the NLO terms. In the infrared, we smoothly freeze coupling to  the value $\as(r\to \infty) = 0.76$.

We parametrize the initial condition for the dipole amplitude as
\begin{equation}
S(r) =
 \exp \left[ 
 -\frac{(r^2 {\qso^2})^{
\gamma
}
 }
 {4} 
\ln \left(\frac{1}{r  {\lqcd}}+ e\right)
\right].
\end{equation}
This form has essentially two tunable parameters, whose values should ultimately be determined by experimental data. The initial saturation scale $\frac{\qso}{\lqcd}$  basically determines value of the coupling $\as$ at the initial rapidity. The other parameter is the anomalous dimension $\gamma$ that affects the shape of the dipole. At LO,  phenomenology prefers values $\gamma\gtrsim 1$ at $y=0$, which then eventually evolve into $\gamma\sim 0.8$ with the evolution.

\begin{figure}
\centerline{\includegraphics[width=0.45\textwidth]{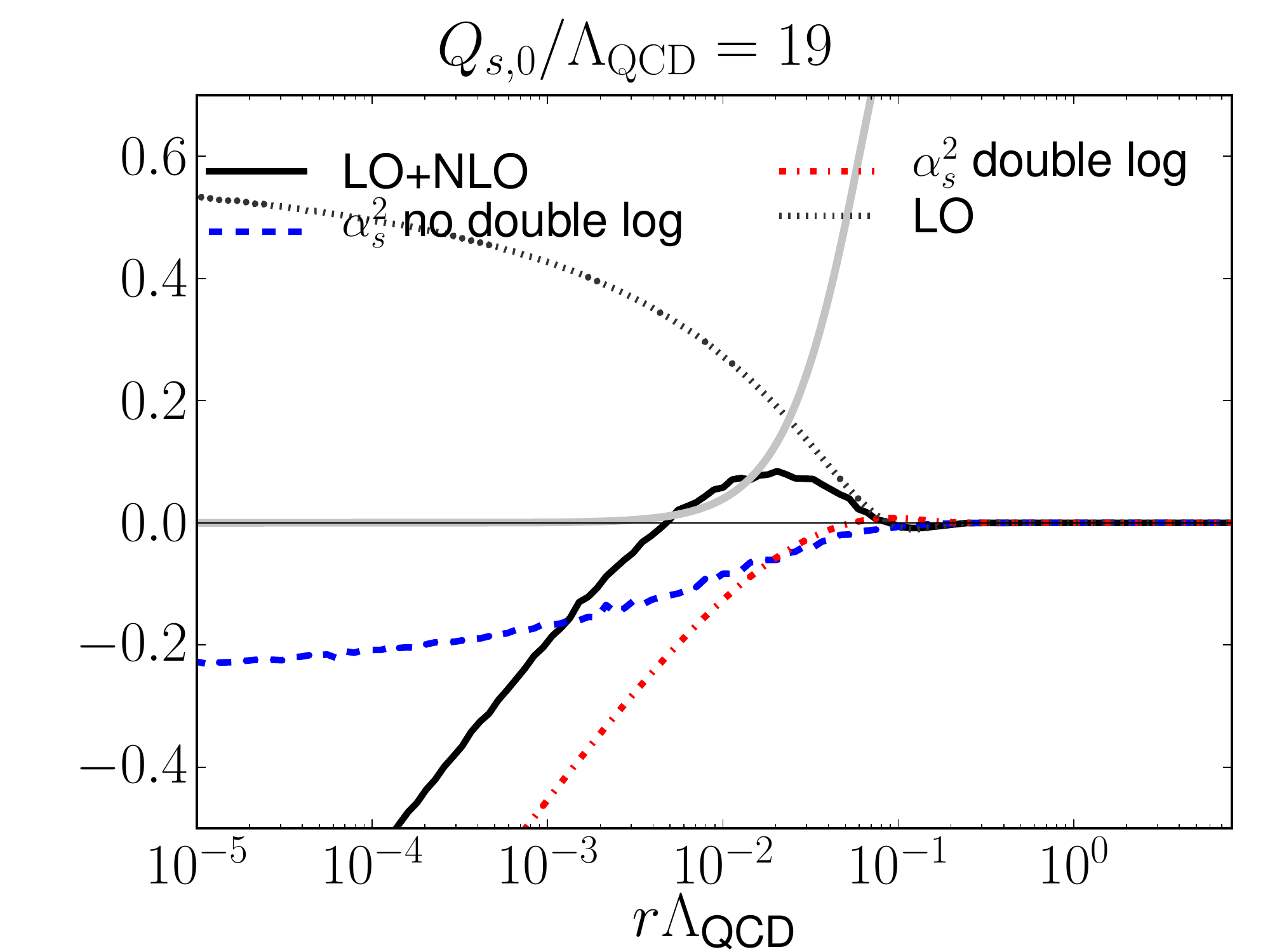}}
\caption{Relative change $\partial_y N(r) / N(r)$ from different terms in the unresummed BK equation \nr{eq:nlobk}.}
\label{fig:unresum}
\end{figure}

\begin{figure*}
\centerline{\includegraphics[width=0.40\textwidth]{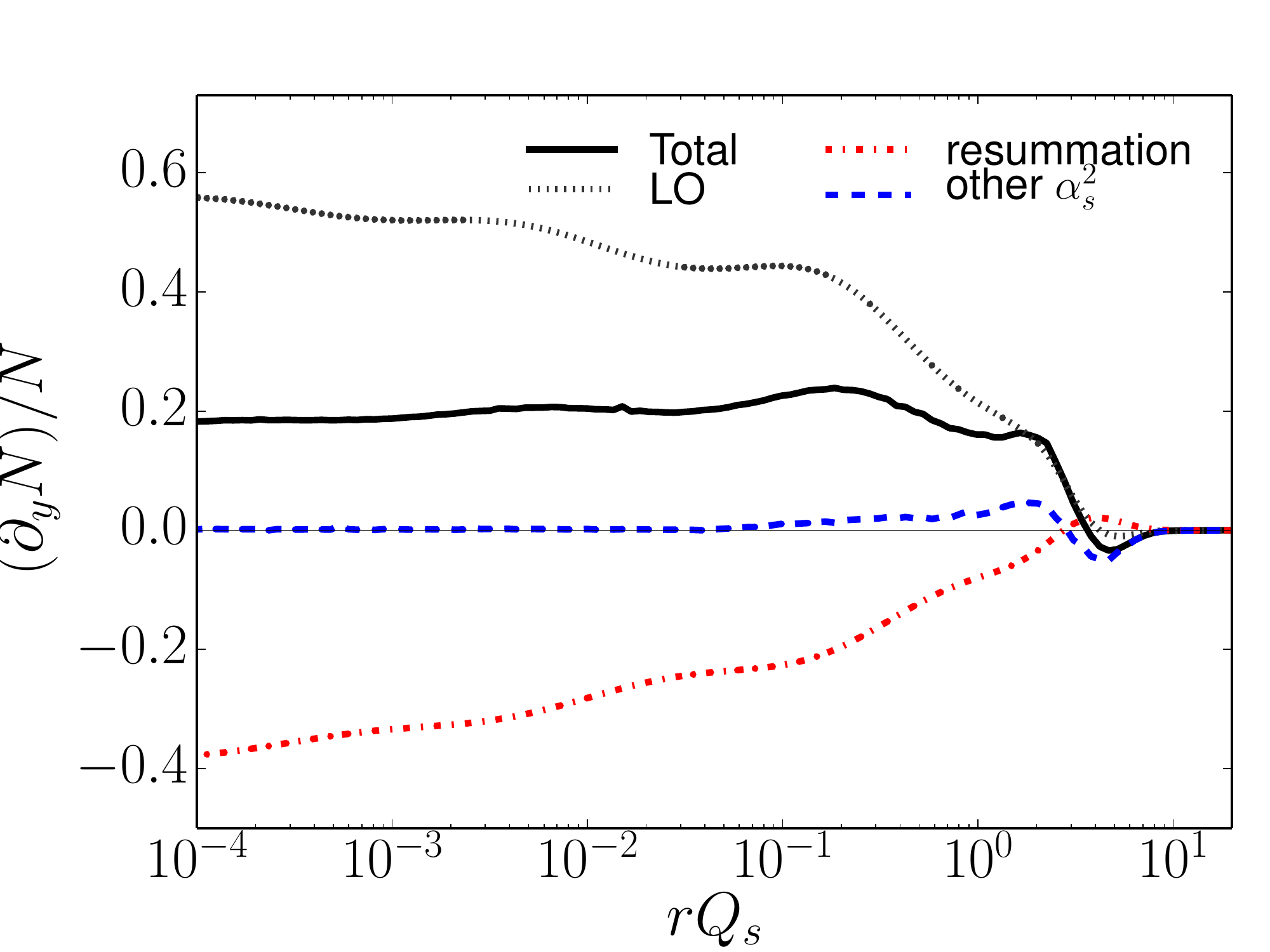}
\rule{1cm}{0pt}
\includegraphics[width=0.40\textwidth]{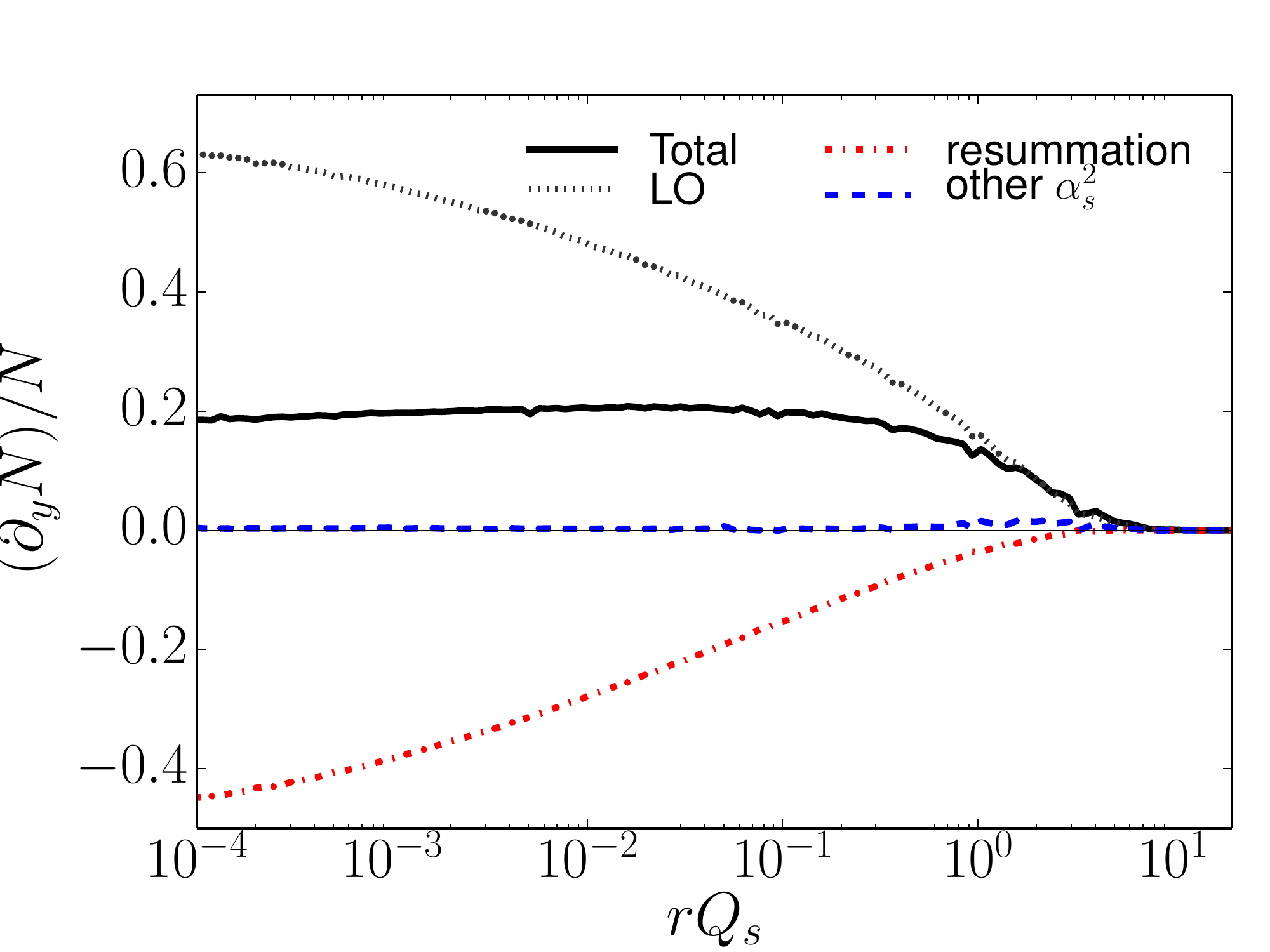}
}
\caption{Contributions of the LO (running coupling), resummation, and other NLO terms to the evolution speed of the amplitude, at the initial rapidity $y=0$ (left) and after $y=10$ units of evolution (right).}
\label{fig:wresum}
\end{figure*}

\section{Double and single log resummation}

Figure \ref{fig:unresum} shows the contribution of different types of terms in \eq\nr{eq:nlobk} to the relative change of the scattering amplitude in one rapidity step, $\partial_y N(r) /N(r)$. It can be seen that the double (nonconformal) logarithm in $K_1$ is responsible for a negative change in the amplitude that diverges logaritmically for small dipole sizes $r$. It was shown in Ref.~\cite{Beuf:2014uia,Iancu:2015vea} that this divergence is related to the derivation of \eq\nr{eq:nlobk} using momentum ordering in the probe. Schematically, the LO and double log terms of the equation can be written in an integral form 
\begin{multline}
S_y(r) = \as \int^y \ud y' \int_z \frac{r^2}{X^2 Y^2}
\left[1+ \frac{\as \nc}{2\pi}  \ln{\frac{X^2}{r^2}} \ln{\frac{Y^2}{r^2}}\right]
\\
\left[ S_{y'}(X)S_{y'}(Y)-S_{y'}(r)\right],
\end{multline}
which is, to this order in $\as$,  equivalent to the ``kinematically constrained'' form
\begin{multline}\label{eq:kc}
S_y(r) = \as \int_z \int^{{\boldsymbol{y-\ln z^2/r^2}}} 
\!\!\!\!\!\!
\ud y' 
\frac{r^2}{X^2 Y^2}
\left[ S_{y'}(X)S_{y'}(Y)-S_{y'}(r)\right].
\end{multline}
The kinematically constrained equation~\nr{eq:kc} resums the problematic double logarithms  and is expected to give a physically meaninful result (see also~\cite{Albacete:2015xza}). As a practical tool it has the disadvantage of corresponding, written in differential form in $y$, to an equation that is nonlocal in rapidity. A clever trick proposed in Ref.~\cite{Iancu:2015vea} allows one to rewrite it in a rapidity-local differential form, but  with the replacement
\begin{equation}
\left[1+ \frac{\as \nc}{2\pi}  \ln{\frac{X^2}{r^2}} \ln{\frac{Y^2}{r^2}}\right]
\to
\frac{
J_1\left( 2\sqrt{ \frac{\as \nc}{\pi}  \ln\frac{X^2}{r^2} \ln\frac{Y^2}{r^2}  } \right)
}{
\sqrt{
 \frac{\as \nc}{\pi}  \ln\frac{X^2}{r^2} \ln\frac{Y^2}{r^2} 
}
}.
\end{equation}

Hidden in the kernel $K_2$ there is also a single transverse logarithm~\cite{Iancu:2015joa} that corresponds to the LO DGLAP  anomalous dimension  $A_1 = 11/12$:
\begin{multline}
\as^2
K_2\otimes
\left[
S(X)S(z-z')S(Y')-S(X)S(Y)
\right]
\\
\sim 
\as^2 A_1 \int_{z} \frac{r^2}{X^2 Y^2} \ln \frac{\min\{X^2,Y^2\}}{r^2}
+ \dots 
\end{multline}
Physically this should be resummed into 
\begin{equation}
K_{STL} = \exp \left\{ - \frac{\as \nc A_1}{\pi} \left|\ln \frac{C_\textnormal{sub} r^2}  {\min\{X^2,Y^2\}}  \right|\right\}.
\end{equation}
The numerical solution in Ref.~\cite{Iancu:2015joa} includes only these resummed double and single logarithms, setting by hand  $C_\textnormal{sub}=1$ and neglecting the other ``finite'' NLO terms. 

\section{Resummed equation}

In Ref.~\cite{Lappi:2016fmu} we set out to remove this limitation and include both the full finite NLO terms and the resummed single and double logarithms. To do this one must remove from the finite NLO part the single log that is already included in $K_{STL}$ to avoid double counting, and render the equation formally independent on the constant $C_\textnormal{sub}$ up to order $\as^2$. We write the full resummed equation as
\begin{multline}
	\partial_y S(r) = \frac{\as \nc}{2\pi^2} 
\left[ K_{DLA}K_{STL}K_{Bal} -K_\textnormal{sub} + K_1^\textnormal{fin}\right]
\\
\otimes [S(X)S(Y)-S(r)] 
\\ 
		+ \frac{\as^2 \nc^2}{8\pi^4} K_2 \otimes [S(X)S(z-z')S(Y')-S(X)S(Y)]
+ \nf \textnormal{-part}.
\end{multline}
To summarize, in the final resummed equation
\begin{itemize}
 \item $K_{DLA} \sim J_1(x)/x$ would give back the double log term in $K_1$ if expanded in $\as$ to order $\as^2$.
\item $K_{Bal}$ resums $\beta$-function terms in $K_1$
\item $K_{STL} = \exp \left\{ - \as \nc A_1 \# \ln r^2\right\}$ resums the single transverse log
\item $K_\textnormal{sub}$ subtracts $\as^2$-part of $\as K_{STL}$, which is already included in $K_2$, to avoid double counting
\item $ K_1^\textnormal{fin}$ contains the other parts of $K_1$ that are not logarithmically enhanced.
\end{itemize}
For illustration purposes we now split this equation into three parts.
\begin{description}
\item[LO] (running coupling) $\frac{\as \nc}{2\pi^2} K_{Bal}$: this includes the terms typically used  in phenomenology so far 
\item[Resummation] $\frac{\as \nc}{2\pi^2} K_{Bal} \left[ K_{DLA}K_{STL}-1\right]$: 
these are the resummed logarithmic terms included in the calculation of Ref.~\cite{Iancu:2015joa}
\item[Finite NLO] includes the other parts  $-K_\textnormal{sub} + K_1^\textnormal{fin}$ and $K_2,K_f$.
\end{description}
The separation between resummation and finite NLO terms depends on the constant $C_\textnormal{sub}$ that cannot be fixed by the analytical calculation of the leading single logarithm. In the following calculation we have numerically determined the value that \emph{minimizes} the ``finite NLO'' term and includes a maximal amount of the evolution in the resummation terms. We find that this can be done by taking  $C_\textnormal{sub}=0.65$.

\begin{figure}
\centerline{
\includegraphics[width=0.45\textwidth]{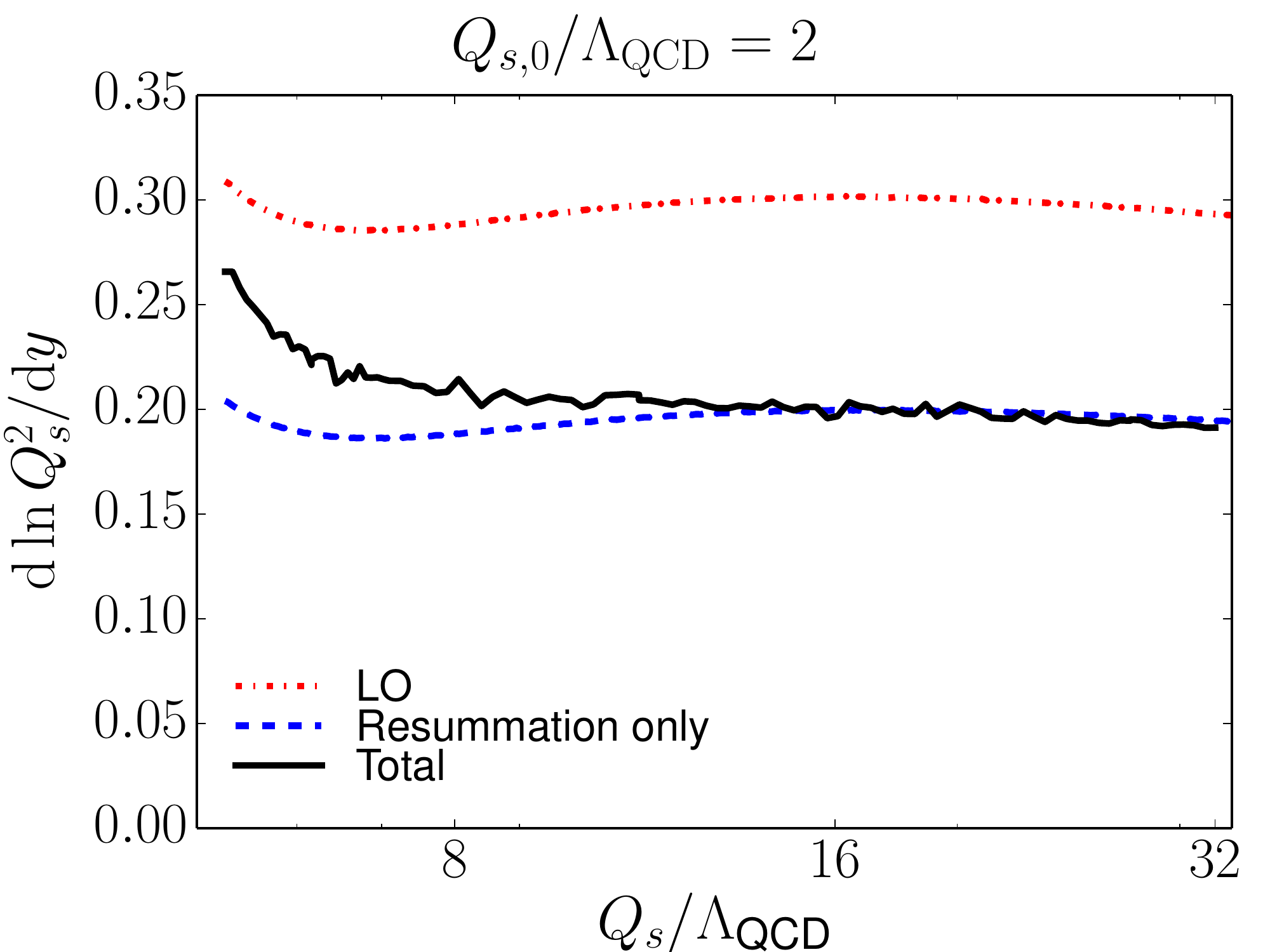}
}
\caption{Evolution speed of the saturation scale.}
\label{fig:lambda}
\end{figure}

Figure~\ref{fig:wresum} shows that the finite NLO terms are a significant contribution to the evolution speed at the initial condition. Asymptotically they become negligible, provided that one has chosen the optimal value for the subtraction constant $C_\textnormal{sub}=0.65$. With an arbitrary value such as $C_\textnormal{sub}=1$ the finite NLO terms would remain significantly larger throughout the phenomenologically interesting rapidity regime. The corresponding evolution speed of the saturation scale, defined as
$\lambda = \ud \ln \qs^2/\ud y$, is shown in Fig.~\ref{fig:lambda}

As a final note that could have a significant effect on phenomenology using the BK equation let us discuss the anomalous dimension, which we define here as
$ \gamma(r) \equiv - \frac{\ud \ln N(r)}{\ud \ln r^2} $ to enable a numerical extraction from the evolved amplitude. In LO running coupling evolution, it quickly approaches a ``geometrical scaling'' value $\gamma\sim 0.8$ independently of the initial condition. The effect of the resummed NLO contributions seems to be to significantly slow down this development: the value of $\gamma(r)$ stays close to the initial condition, at least for initial values of $\gamma$ in the range $0.8 \dots 1.2$. The fate of the geometrical scaling phenomenon in NLO evolution is thus not obvious and deserves more careful study in the future.

\section{Conclusions}

In conclusion, we have performed the first numerical solution of the BK equation that includes both a resummation of problematic transverse logarithms, and the full nonlogarithmic NLO terms. The resummation of the double logarithms stabilizes the equation and enables a meaningful numerical solution. Also the single ``DGLAP'' type logs can be resummed, but this procedure leaves an ambiguity in terms of an undetermined constant under the logarithm which we denote as $C_\textnormal{sub}$. The dependence on this constant cancels to this order in $\as$ when also the other NLO terms are consistently included, removing the double counting between the two. Numerically we find that (at least for initial conditions $\gamma=1$) the value  $C_\textnormal{sub}=0.65$ minimizes the finite NLO terms. Our recommendation would be to use this this value if, for numerical convenience, one desires to work with the simpler resummation terms only. In the future this procedure should be used to fit DIS data to extract a parametrization for the initial condition of the evolution, and to apply the resummed NLO equation to particle production in proton-nucleus and nucleus-nucleus collisions.

T.L.  is supported by the Academy of Finland, projects 
267321, 273464 and 303756 and  by the European Research Council, grant
 ERC-2015-CoG-681707. H.M. is supported under DOE Contract No. {DE-SC0012704}.
We acknowledge computing resources from
CSC -- IT Center for Science in Espoo, Finland.

\bibliographystyle{elsarticle-num}
\bibliography{spires}







\end{document}